\documentclass{aa}  

\usepackage{graphicx}
\usepackage{txfonts}
\usepackage[varg]{newtxmath}
\usepackage{booktabs}
\usepackage{xcolor}
\usepackage{media9}

\definecolor{grey}{rgb}{0.75,0.75,0.75}
\definecolor{orange}{rgb}{1.0,0.5,0.15}
\definecolor{brown}{rgb}{0.7,0.25,0.0}
\definecolor{pink}{rgb}{1.0,0.5,0.5}
\definecolor{darkerred}{rgb}{0.8,0,0}
\definecolor{darkerblue}{rgb}{0,0,0.8}
\definecolor{blue}{rgb}{0,0.08,0.65}
\definecolor{red}{rgb}{0.65,0.08,0.05}
\definecolor{green}{rgb}{0.15,0.45,0.25}

\newcommand{\at}[2][]{#1|_{#2}}

\begin{document}

   \title{Magnetic wind-driven accretion in dwarf novae}
   \author{Nicolas Scepi
		\and
		Guillaume Dubus
		\and
		Geoffroy Lesur
          }

   \institute{
   	Univ. Grenoble Alpes, CNRS, IPAG, 38000 Grenoble, France 
	     }

   \date{Received ; accepted ; in original form \today}

 
  \abstract
   {Dwarf novae (DNe) and X-ray binaries exhibit outbursts thought to be due to a thermal-viscous instability in the accretion disk. The disk instability model (DIM) assumes that accretion is driven by turbulent transport, customarily attributed to the magneto-rotational instability (MRI). However, recent results point out that MRI turbulence alone fails to reproduce the light curves of DNe.}
   {Our aim is to study the impact of wind-driven accretion on the light curves of DNe. Local and global simulations show that magneto-hydrodynamic winds are present when a magnetic field threads the disk, even for relatively high ratios of thermal pressure to magnetic pressure ($\beta \approx 10^{5}$). These winds are very efficient in removing angular momentum but do not heat the disk, thus they do not behave as MRI-driven turbulence.}
   {We add the effect of wind-driven magnetic braking in the angular momentum equation of the DIM but neglect the mass loss due to the wind. We assume a fixed magnetic configuration: dipolar or constant with radius. We use prescriptions for the wind torque and the turbulent torque derived from shearing box simulations.}
   {The wind torque enhances the accretion of matter, resulting in light curves that look like DNe outbursts when assuming a dipolar field with a moment $\mu\approx10^{30}\,\mathrm{G\,cm^{3}}$.  In the region where the wind torque dominates the disk is cold and optically thin, and the accretion speed is super-sonic. The inner disk behaves as if truncated, leading to higher quiescent X-ray luminosities from the white dwarf  boundary layer than expected with the standard DIM. The disk is stabilized if the wind-dominated region is large enough, potentially leading to ``dark'' disks that emitting little radiation.}
   {Wind-driven accretion can play a key role in shaping the light curves of DNe and X-ray binaries. Future studies will need to include the time evolution of the magnetic field threading the disk to fully assess its impact on the dynamics of the accretion flow.}

   \maketitle
%

\section{Introduction}
Dwarf novae (DNe) are binaries composed of a white dwarf accreting via Roche-lobe overflow from a low-mass stellar companion. Dwarf novae are known for their optical outbursts, with the prototypical U Gem-like outburst lasting a week and recurring monthly. These outbursts originate in the accretion disk surrounding the white dwarf. The thermal instability responsible for the outbursts is due to the strong  temperature dependence of the opacity when hydrogen is being ionized. In this regime the disk is thermally unstable and has to reach a colder or hotter temperature to stabilize in the quiescent or the eruptive state, respectively. This is the core of the disk instability model (DIM; \citealt{lasota2001}), which aims to explain the outbursts of DNe and X-ray binaries (where the compact object is a black hole or a neutron star). 

The shape of the light curves obtained with the DIM depends on the mass accretion rate in the eruptive and quiescent states. In the DIM, accretion is only driven by turbulence, with the turbulence usually attributed to the magneto-rotational instability (MRI; \citealt{balbus1991}). In this case, the angular momentum transport and heating are both parametrized by the well-known parameter $\alpha$ \citep{Shakura:1973vo}. This leads to the long-standing problem that  $\alpha\approx0.1$ is required in eruption, \citep{kotko2012} whereas $\alpha\approx0.03$ is required in quiescence (\citealt{cannizzo1988}, \citealt{cannizzo2012}). There has been extensive work to know if MRI transport could provide such a dichotomy (\citealt{latter2012}, \citealt{2014ApJ...787....1H}). Recent results indicate that MRI turbulence alone fails to reproduce the light curves of DNe. A disk instability model using $\alpha$ from MRI simulations does not reproduce well the outburst amplitudes and timescales \citep{2016MNRAS.462.3710C}. In particular, the eruptions have ``reflares'' that are not observed. Moreover, in the quiescent state, MRI turbulence is not able to sustain angular momentum transport due to resistivity (\citealt{gammie1998}, \citealt{2018A&A...609A..77S}).

Until now, every model of DNe has used an $\alpha$ parametrization, presupposing that transport of angular momentum can be modeled as an effective viscosity. However, magneto-hydrodynamic (MHD) outflows extract angular momentum very efficiently and cannot be reduced to an $\alpha$ prescription. It has long been thought that MHD outflows require a strong magnetization, {\em i.e.} $\beta\approx1$, where $\beta$ is the ratio of thermal to magnetic pressure. In this case, the magnetic field collimates the outflow in what is usually called a jet \citep{ferreira1995}. It is now becoming clear from both local and global simulations that winds due to MRI can be launched even for $\beta\approx10^{2}$ to $10^{5}$ (\citealt{fromang2013}, \citealt{lesur2013}, \citealt{bai2013a}, \citealt{2018arXiv180909131S}, \citealt{zhu2018}). The basic mechanism that allows the formation of winds for such low magnetizations is presented in \cite{lesur2013}, \cite{fromang2013} and \cite{bai2013a}, and can be understood as follows. Because of the MRI amplification mechanism, a large amount of toroidal field is created from the poloidal field in the upper layers of the disk. The resulting vertical gradient of magnetic pressure accelerates matter vertically to launch an outflow, at the surface where $\beta\approx1$, while it compresses the disk at the same time. Angular momentum is extracted from the disk, stored in the toroidal field, and is ultimately given back to outflowing material. From this scenario, we see that winds or jets are unavoidable in an MRI-unstable disk and in the presence of a large-scale poloidal field, even for midplane magnetization much lower than 1. This calls into question the hypothesis of a purely turbulent disk, which has been used extensively to model eruptions of DNe.

Winds are seen during eruptions of DNe (\citealt{cordova1982}, \citealt{mauche1987}). Mass loss rates are believed to be a small percentage of the mass accretion rate (\citealt{hoare1993}, \citealt{knigge1997}) and their velocities correspond to the escape velocity from the inner regions of the disk \citep{cordova1982}. There may be some indirect evidence for winds in quiescence \citep{santisteban2018}. However, the lack of direct evidence for winds in quiescence does not mean that they are absent in this state, as they are expected to be more difficult to see than in eruption \citep{drew1990}. The mechanisms of acceleration of the observed winds are still unclear, yet much of the previous studies of winds in eruptions of DNe have focused on line-driven winds (\citealt{proga1998}, \citealt{feldmeier1999}, \citealt{knigge1999}, \citealt{drew2000}, \citealt{proga2002}). There has also been some work to characterize the magneto-centrifugal effects on line-driven winds (\citealt{proga2000}, \citealt{proga2003}) or on the radiative signatures in stationary disks \citep{knigge1999}. However, none of these studies has considered the dynamical impact of MHD winds that could also be present in quiescence, and in particular the impact of radial accretion driven by the effective magnetic braking of these winds.

Recently, \cite{2018arXiv180909131S} proposed, based on the results of shearing box simulations threaded by a large-scale magnetic field, that MHD winds could be very important in driving accretion in DNe, especially in the quiescent state, which is easily dominated by the magnetic field. New solutions may appear where the mass accretion rate is high but the density is lower than in a standard accretion disk, similar to the jet-emitting disk solution of \citet{ferreira1995}. In this paper, we include the effect of a wind in the disk instability model by including only the magnetic wind torque and neglecting the mass lost in the wind. We explore the effects on the light curves and stability of DNe.

\section{Model for the disk evolution}
We use the disk evolution code of \citet{1998MNRAS.298.1048H}, modified to take the wind torque into account. This code solves the system of time-dependent, radial differential equations obtained after assuming specific forms for the angular momentum transport and energy release. Specifically, we use an $\alpha$ prescription to describe the turbulent MRI-driven angular momentum transport and heating, which is appropriate for thin disks \citep{Balbus:1999mi}. MRI is therefore included as a viscous term in the angular momentum and energy equations. 

The first difference is that instead of taking a constant $\alpha$, as was done by \citet{1998MNRAS.298.1048H}, we use the functional form of $\alpha$ derived from our shearing box simulations \citep{2018arXiv180909131S}
\begin{equation}
\frac{\alpha}{0.01}=f \left[3.65+5.35 e^{-x^{2}/2}+0.165\tanh x+\frac{137}{\beta^{0.58}}\right]
\label{alpha}
\end{equation}
with
\begin{equation}
x\equiv\left(\frac{T_{\rm eff}-6866\rm\,K}{853\rm\,K}\right)
\end{equation}
where $T_{\rm eff}$ is the local effective temperature of the disk, and with $f$ defined as
\begin{equation}
f\equiv\frac{\beta}{1+\beta}.
\label{eq:f}
\end{equation}
Here $\beta$ is the plasma parameter
 \begin{equation}
\beta \equiv \frac{8\pi P}{B_z^2}
\end{equation}
with $P$ the midplane gas pressure (radiation pressure is negligible), and $B_z$ the vertical component of the magnetic field, which sets the net flux in the MRI shearing box simulations.
The code solves vertically integrated equations and relies on a set of pre-calculated vertical structures to get the radiative cooling term $Q^-=2\sigma T_{\rm eff}^4$ for a given radius $r$, surface density $\Sigma$ and midplane temperature $T_c$. These vertical structures were calculated as in \citet{1998MNRAS.298.1048H}, assuming that the vertical structure is optically thick and including convection in the mixing length approximation. We did not attempt to take into account the differences in vertical profiles between these 1D calculations and MRI shearing box simulations in the presence of strong convection or a strong net magnetic flux \citep{2016MNRAS.462.3710C,2018arXiv180909131S}. This would require a more realistic profile of $\alpha(z)$ given by MRI. However, given the wide range of magnetization values used in our simulations, this is not a straightforward task.

The second difference is that we include a term to account for wind-driven angular momentum transport. This is most clearly expressed by writing that the mass accretion rate $\dot{M}$ at a radius $r$ is given by \citep{2018arXiv180909131S}
\begin{eqnarray}
\dot{M}&=&6\pi r^{1/2}\frac{\partial}{\partial r} \left( \nu \Sigma r^{1/2} \right)+\frac{r}{\Omega}q B_{z}^2\\
&\equiv&\dot{M}_{r\phi}+\dot{M}_{z\phi}
\label{eq:momentum}
\end{eqnarray}
where $\Omega$ is the Keplerian angular velocity and $q$ is defined below (Eq.~\ref{qbeta}). This expression also holds for time-dependent disks. $\dot{M}_{r\phi}$ is the mass flow rate due to the viscous torque and $\dot{M}_{z\phi}$ is that due to the magnetic wind torque. We note that these are not mass loss terms. In fact, we do not include mass loss due to the wind in the continuity equation. The shearing box simulations only give an upper limit on the magnitude of the wind mass loss \citep{2018arXiv180909131S}, whereas global simulations from \cite{zhu2018} typically indicate $\sim 1\%$ of the local $\dot{M}$ is lost to the wind. We assumed here, for simplicity, that the angular momentum is carried away in the wind by a negligible amount of mass, so the mass conservation remains unchanged. In the presence of turbulence alone, all the gravitational energy extracted by the radial stress ends up heating the disk. This leads to the famous relation between the heating rate $Q^+$ and the radial stress tensor $W_\mathrm{r\phi}$: 
\begin{equation}\label{eq:heating}
\begin{aligned}
Q^+&\equiv\frac{3}{4}\Sigma\Omega W_\mathrm{r\phi} \\
&\equiv\frac{3}{4}\Sigma\Omega\alpha P
\end{aligned}
\end{equation}
There is no heating associated with the wind torque itself \citep{2018arXiv180909131S}. However, a fraction of the extracted accretion energy can be lost in the wind, making the disk cooler than it would be if it were only viscously accreting.  In the following, we neglect this effect, assuming that the wind has no direct impact on the energy budget of the disk. The DIM energy equation therefore remains unchanged; the heating rate due to turbulence is given by Eq.\,\ref{eq:heating}.
 
The magnitude of the wind torque is parametrized by $q$, whose functional dependence on $\beta$ was constrained by the MRI shearing box simulations as
\begin{equation}
q=f\left(8\times 10^{-5} \beta^2+1.8\times 10^5\right)^{0.3}
\label{qbeta}
\end{equation}
Compared to the expressions in \citet{2018arXiv180909131S},  $q$  (Eq.~\ref{qbeta})  and $\alpha$ (Eq.~\ref{alpha}) have both been multiplied by $f$ (Eq.~\ref{eq:f}). This ensures that both the viscous and wind torques drop to 0 when $\beta$ drops below 1. In this case, MRI transport ceases and Rayleigh-Taylor instabilities, such as the interchange instability, will take over angular momentum transport \citep{mckinney2012}. In a real disk, the magnetic field diffusion and the interchange instability may act to prevent $\beta$ from falling much lower than 1. A better treatment of these very magnetized regions is beyond the scope of this paper. In our simple model, the lower limit on $\beta$ is set by the specific form of $f$.

The third and final difference with \citet{1998MNRAS.298.1048H} is that, instead of the no (viscous) torque inner boundary, we assume
\begin{equation}
\dot{M}_{\rm in}=\left(3\pi \nu \Sigma+\dot{M}_{z\phi}\right)_{\rm in}
\label{eq:boundary}
\end{equation}
at the inner edge. This makes no difference to the light curves, but it is computationally easier when the wind torque dominates the inner disk. At the outer radius, we set $\dot{M}=\dot{M}_t$, the mass transfer rate from the companion. We chose to fix the outer disk radius $R_d$, ignoring the interaction between the wind, viscous, and tidal torques that remove the disk angular momentum at the outer edge. The influence on the light curves of a varying outer radius is minor compared to the effects we explore here. 

We used two different prescriptions for $B_z$, which we assumed constant in time: a constant $B_z$ with radius and a dipolar form $B_z=\mu r^{-3}$, with $\mu$ the dipole moment. We take these two prescriptions to roughly represent what might be expected if the magnetic field of the companion and the white dwarf, respectively, is the source of the net magnetic flux threading the disk. In the case of a dipolar field, we assume that the field is disconnected from the white dwarf, so that the wind torque is always positive in equation \ref{eq:momentum} (i.e., we do not allow for propeller regimes).

The true distribution of $B_z$  is a complex function of radius and time, depending upon the source of the field, its advection, and its diffusion or reconnection in the disk (Sect.\,\ref{sec:disc}). Including these complex effects is beyond the scope of this study and will the subject of future work.

To summarize, a disk model is therefore fully set by the mass of the white dwarf $M$, the outer disk radius $R_d$, the mass transfer rate $\dot{M}_t$, and the chosen distribution of $B_z$. The inner disk radius is the white dwarf radius, directly related to $M$. In the following, we assume $M=0.6\rm\,M_\odot$, which implies an inner disk radius $R_{\rm in}\approx 8.7\times 10^8\rm\,cm$, and $R_d=2\times 10^{10}\rm\,cm$. We write $\mu_{30}=\mu/(10^{30}\rm\, G\,cm^3)$.

\section{Steady, stable disks}
\begin{figure}
\begin{center}
\includegraphics[width=\linewidth]{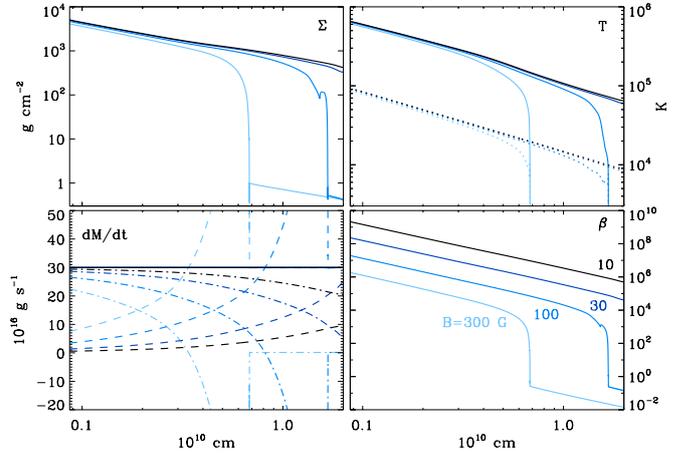} 
\caption{Radial structure of steady, stable disks with constant $B_z$. Clockwise from top left: surface density $\Sigma$ ; midplane temperature $T_c$ (solid) and effective temperature $T_{\rm eff}$ (dotted); plasma $\beta$; mass flow rate $\dot{M}$ (solid), $\dot{M}_{r\phi}$ (dot-dashed line), and $\dot{M}_{z\phi}$ (dashed line). Here,  $\dot{M}=\dot{M}_{t}=3\times10^{17}\rm\,g\,s^{-1}$. Four radial structures are shown corresponding to $B_{z}=10, 30, 100, 300\rm\,G$ (see bottom right panel for color-coding). }
\label{fig:stabhotcomp}
\end{center}
\end{figure}
\begin{figure}
\begin{center}
\includegraphics[width=\linewidth]{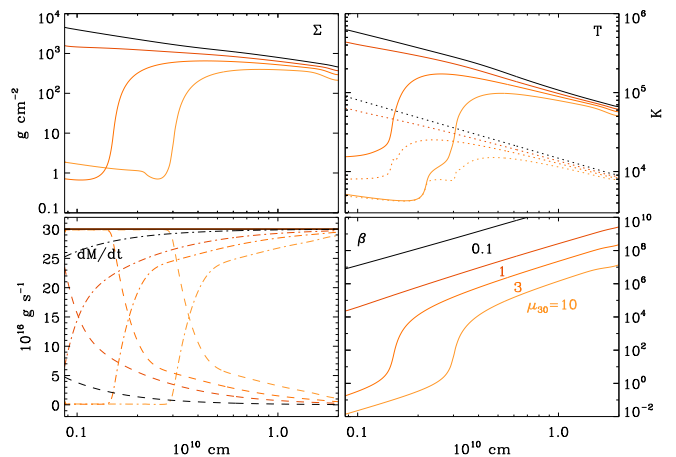} 
\caption{Same as Figure~\ref{fig:stabhotcomp} with $B_z=\mu r^{-3}$. Four radial structures are shown corresponding to $\mu=\left[0.1,1,3,10\right]\times 10^{30}\rm\,G\,cm^{3}$ (see bottom right panel for colorcoding).}
\label{fig:stabhotdipole}
\end{center}
\end{figure}

\subsection{Radial structure including a wind torque}
Figures \ref{fig:stabhotcomp} and \ref{fig:stabhotdipole} show how the disk radial structure changes with increasing wind torque, assuming a constant (over $r$) or dipolar $B_z$. The reference radial structure in both figures is a steady, hot disk with $\dot{M}_t=3\times 10^{17}\rm\,g\,s^{-1}$. Such a disk is stable against the thermal-viscous instability since it is everywhere hot enough for hydrogen to be ionized (see \S\,\ref{sec:stab}). The pressure drops as $P\propto r^{-21/8}$ in an unmagnetized, viscous steady disk \citep{Shakura:1973vo} so that $\beta \propto  r^{-21/8}$ decreases with $r$ when $B_z$ is constant, or $\beta\propto r^{27/8}$ increases with $r$ when $B_z$ is dipolar. Increasing $B_z$ therefore influences first the outer disk when $B_z$ is constant, or the inner disk first when $B_z$ is dipolar. 

Both figures show that $\Sigma$ and $T_c$ decrease substantially in the regions of decreasing $\beta$. The decrease in $\Sigma$ is  due to an increase in the radial velocity $v_r$ of the material when the wind torque becomes important,  the mass accretion rate $\dot{M}=-2 \pi R \Sigma v_r$ being constant in the steady disk. Equation \ref{eq:momentum} shows that $v_r \approx -2 \left(q/\beta\right) c_s$ if $\dot{M}\sim \dot{M}_{z\phi}$. The wind-driven accretion velocity is super-sonic, much faster than the standard viscous radial accretion $v_r\approx- \alpha (h/r) c_s$ (with $h$ the disk height and $h/r\ll1$). The decrease in $\Sigma$ signals efficient transport of angular momentum by the wind torque. It does not signal mass loss to the wind, which is not taken into account here (but would  add to the effect). The decrease in $T_c$ is directly linked to the reduced heating when the fraction of the mass accretion due to the viscous torque diminishes.

The lower left panels of Figures \ref{fig:stabhotcomp} and \ref{fig:stabhotdipole} show that $\dot{M}_{z\phi}$ and $\dot{M}_{r\phi}$ have very different behaviors in the constant and dipolar $B_z$ cases. When $\beta\gg 1$, $\dot{M}_{r\phi}$ is constant with radius and equal to $\dot{M}$, while the wind flow rate varies with $r$. In the constant $B_z$ case, $\dot{M}_{z\phi}\propto r^{37/40}$ when $\beta\gg 1$, whereas $\dot{M}_{z\phi}\propto r^{-59/40}$ in the dipolar case. In the latter case, $\dot{M}_{z\phi}$ is stronger close to the white dwarf and the density $\Sigma$ adjusts itself to have $\dot{M}\approx \dot{M}_{z\phi}$ at the inner edge (see Sect.\,\ref{sec:wd} below). The  decrease in $\dot{M}_{z\phi}$ with $r$ ensures that $\dot{M}_{r\phi}$ and $\dot{M}_{z\phi}$ are $\leq\dot{M}$ everywhere (Eq.\,\ref{eq:momentum}) with $\dot{M}\approx \dot{M}_{r\phi}$ at the outer radius. However, with a constant $B_z$, $\dot{M}_{r\phi}$ dominates close to the white dwarf and the increase in the wind torque with $r$ must be checked by $\dot{M}_{r\phi}$ to keep $\dot{M}$ constant (Eq.\,\ref{eq:momentum}). Whereas $\dot{M}_{z\phi}$ is necessarily positive, $\dot{M}_{r\phi}$ can take on negative values to accommodate this. Indeed,  $\dot{M}_{z\phi}$ and $|\dot{M}_{r\phi}|$ both become much greater than $\dot{M}$, with the high inflow rate due to the wind torque compensated by a high disk outflow rate from the viscous torque (negative values of $\dot{M}_{r\phi}$).

\subsection{Wind-dominated regions\label{sec:wd}}
The radial profiles change abruptly where $\beta \la 10^4$ {\em i.e.} where $q$ becomes nearly constant (Eq.\,\ref{qbeta}). At this stage, $|v_r|\propto 1/\beta$ so that any increase in $v_r$ results in a decrease in $\Sigma$, and thus in $\beta$, leading to an increase in $v_r$. A runaway decrease in density and temperature occurs until a floor is found at $\beta\approx 1$, when the factor $f$ acts to lower the wind torque. Thus, this floor depends completely on our assumption for $f$. In reality, the value of $\Sigma$ depends on the details of how MRI and the wind torque are gradually quenched as $\beta$ approaches 1. In addition, the temperature drops since viscous heating becomes negligible so that the disk becomes optically thin, stretching our assumptions for the radiative transfer. Despite these caveats, it is clear that, due to the runaway, the values of $\Sigma$ and $T_c$ will be much lower where $\beta\la 10^4$ than in the rest of the disk .

A surprising result is that the disk is stable to the thermal-viscous instability, even though the disk transitions to a low temperature in the wind-dominated regions, passing through the region of strong opacity change due to hydrogen recombination. We observe in the transiting region that the radial advection of thermal energy is non-negligible compared to the heating rate. The thermal equilibrium stability criterion which is usually written as 
\begin{equation}
\frac{\partial Q^+}{\partial T} \at[\bigg]{\Sigma}<\frac{\partial Q^-}{\partial T} \at[\bigg]{\Sigma}
\end{equation}
is known to be invalid in this case. It seems that where wind-driven accretion is dominant the thermal energy is advected at super-sonic velocity preventing the thermal instability from growing. We cross-checked the numerical stability of the steady state solutions using a different code that does not include radial advection in the advection of energy. This code solves the usual mass and angular momentum conservation equations in a way similar to the full disk evolution code of \citet{1998MNRAS.298.1048H}, but it uses a simplified vertical structure with power-law opacities, following Appendix A of \citet{latter2012}. In addition, the energy equation only takes into account viscous heating and radiative cooling, excluding radial energy transport terms which are present in the full disk evolution code. In this case, we find quantitatively similar behaviors to the one presented here, except for very small oscillations of the transition radius around the equilibrium value showing that the disk becomes thermally unstable again. The thermal instability, however, is confined to a very narrow region between the two thermally stable zones. The accretion timescale is too short in the wind-dominated region to enable the formation of thermal-viscous heat fronts. 

The wind-dominated regions with $\beta\la 10^4$ have high accretion velocities and low densities, and are optically thin.  They are essentially dark since the wind-torque does not release heat in the disk \citep{2018arXiv180909131S}. Their observational impact in the optical is identical to truncating the outer disk for a constant $B_z$, or to truncating the inner disk in the dipole case. We find that the location of the wind-dominated region can be estimated analytically by finding the radius where 
\begin{equation}
\dot{M}\approx \xi \dot{M}_{z\phi}(\beta=1)
\end{equation}
where $\xi$ is a constant. For $B_z$ constant, we found $\xi=1/4$ adjusts the numerical results well, giving a wind-dominated radius
\begin{equation}
r_{B}=1.7\times 10^{11}\, B_1^{-4/5} \dot{M}_{16}^{2/5} M_1^{1/5}\rm\,cm
\label{eq:rb}
\end{equation}
For a dipolar $B_z$, $\xi=1$ works well, so that
\begin{equation}
r_{\mu}=1.9\times 10^{9}\, \mu_{30}^{4/7} \dot{M}_{16}^{-2/7} M_1^{-1/7}\rm\,cm
\label{eq:rmu}
\end{equation}
with $B_1=B/1\rm\,G$, $M_1=M/M_\odot$ and $\dot{M}_{16}=\dot{M}/10^{16}\rm\,g\,s^{-1}$. For example, these formulas give $r_B\approx 6.3\times 10^{9}\rm\,cm$ for $B=300\rm\,G$ (Fig.~\ref{fig:stabhotcomp}) and $r_\mu\approx 3.0\times 10^{9}\rm\,cm$ for $\mu_{30}=10$ (Fig.~\ref{fig:stabhotdipole}), in good agreement with the numerical results. 

In the dipole case, $r_\mu$ can be compared to the Alfv\'en radius \citep{Frank:2002fo}
\begin{equation}
r_{\rm A}\approx\left(\frac{\mu^4 }{GM\dot{M}_{\rm in}^2} \right)^{1/7}\approx 6.9\times 10^{8}\,\mu_{30}^{4/7} \dot{M}_{16}^{-2/7}M_1^{-1/7} \rm\,cm,
\label{eq:trunc}
\end{equation}
which sets the radius where matter accretes along the dipolar field lines, {\em i.e.}, the  magnetospheric truncation of the disk. Unsurprisingly, $r_{\mu}$ and $r_A$ have  the same dependencies on parameters since both depend on setting $\beta=1$. Importantly, the Alfv\'en radius is always smaller than the radius $r_\mu$ at which the inner disk becomes wind-dominated. 

\section{DIM-unstable disks}

\subsection{Light curves}
\begin{figure}
\begin{center}
\includegraphics[width=\linewidth]{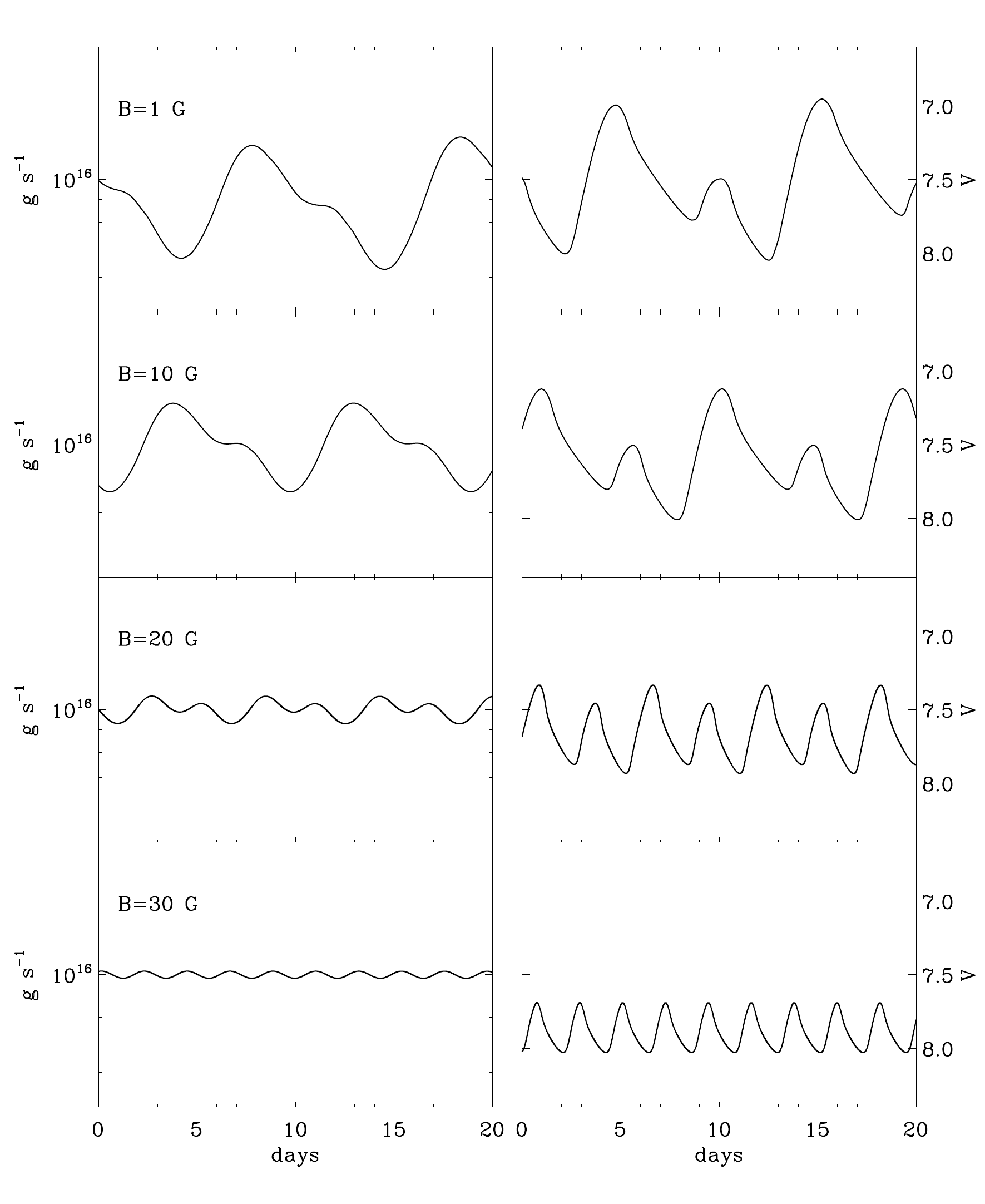} 
\caption{Light curves of unstable disks assuming, from top to bottom, a constant $B_z=1,10, 20, 30\rm\,G$ and $\dot{M}_{\rm t}=10^{16}\rm\,g\,s^{-1}$. Left panels : Mass accretion rate at the inner radius. Right panels : Absolute $V$ magnitude calculated as in \citet{2018A&A...617A..26D}.}
\label{fig:lccomp}
\end{center}
\end{figure}
\begin{figure}
\begin{center}
\includegraphics[width=\linewidth]{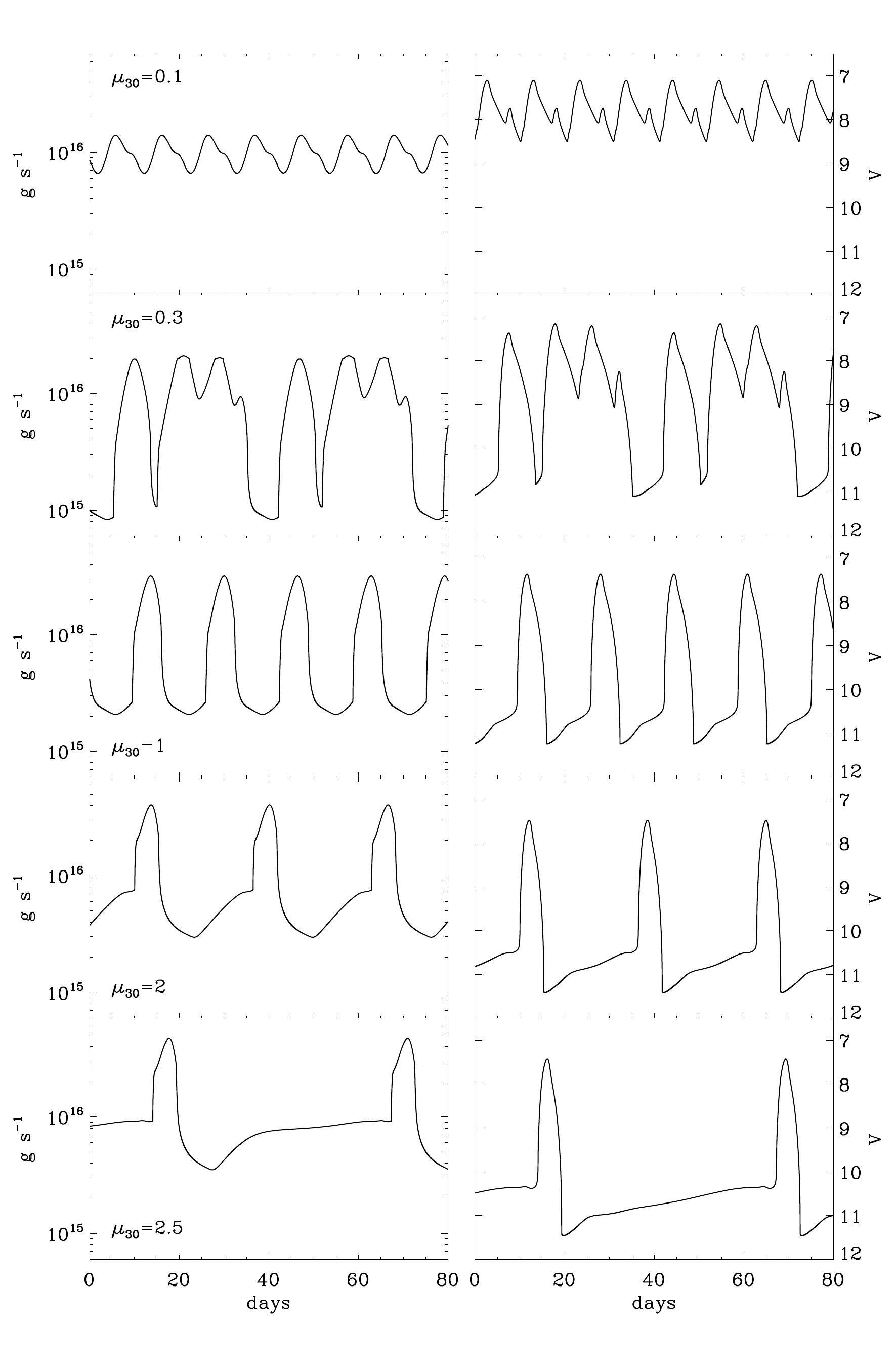} 
\caption{Same as Figure~\ref{fig:lccomp}, but assuming a dipole magnetic field distribution for $B_z$ with (from top to bottom) $\mu=\left[0.1,0.3,1,2,2.5\right]\times 10^{30}\rm\,G\,cm^{3}$.}
\label{fig:lcdipole}
\end{center}
\end{figure}
Figures \ref{fig:lccomp} and \ref{fig:lcdipole} illustrate how the light curves change as we vary the strength of the magnetic field. The disk has $\dot{M}_{t}=10^{16}\rm\,g\,s^{-1}$ so that it is unstable to the thermal-viscous instability in the absence of the wind torque. 

Starting with a steady solution, the disk becomes unstable near the transition between the hot ionized inner region and the cold nearly neutral outer region. Heat fronts then propagate through the part of the disk which is not dominated by the wind, as it cycles through the hot and cold states. We find that these fronts do not propagate far when $B_z$ is low, leading to outbursts  of weak amplitude and with a short recurrence timescale (top panels of Figs.\,\ref{fig:lccomp} and \ref{fig:lcdipole}). Here, the amplitude and recurrence of the cycle depend critically on having a much higher $\alpha$ in outburst than in quiescence. When $\beta\gg 1$, $\alpha$ is nearly constant during the cycle except for an increase close to the tip of the hot branch at $T_{\rm eff}\approx 7\,000\rm\,K$ due to convection (Eq.\,\ref{alpha}, \citealt{2014ApJ...787....1H}). However, this increase is insufficient to produce light curves that look like dwarf novae light curves, as previously found by \citet{2016MNRAS.462.3710C}.

The discrepancy worsens with increasing (constant) $B_z$. The lower panels of Fig.\,\ref{fig:lccomp} show weakening outburst amplitudes and decreasing periodicity. We attribute this to the decrease in the amount of mass available in the outer disk as the increased wind torque lowers $\Sigma$, and to the effect of the wind-dominated region, which acts like a truncation of the outer disk. The average $V$ magnitude increases as less of the disk is hot enough to contribute to the optical flux. A decreasing outer disk radius $R_{\rm d}$ would lead to very similar light curves. At $B\ga 50\rm\,G$, the disk actually becomes stable (staying in `outburst') as  the wind-dominated region has now in effect truncated the viscous, cold outer region. The radial structure then becomes similar to those shown in Fig.\,\ref{fig:stabhotcomp} with a hot, viscous inner disk and a cold, wind-dominated outer disk. The wind torque can thus stabilize a disk that is a priori unstable to the thermal-viscous instability.

\begin{figure*}
\centering\includemedia[
      activate=onclick,
      width=300pt,height=200pt,
      addresource=disk.mp4,
      flashvars={source=disk.mp4 & loop=true & showinfo=0}  
   ]{\includegraphics{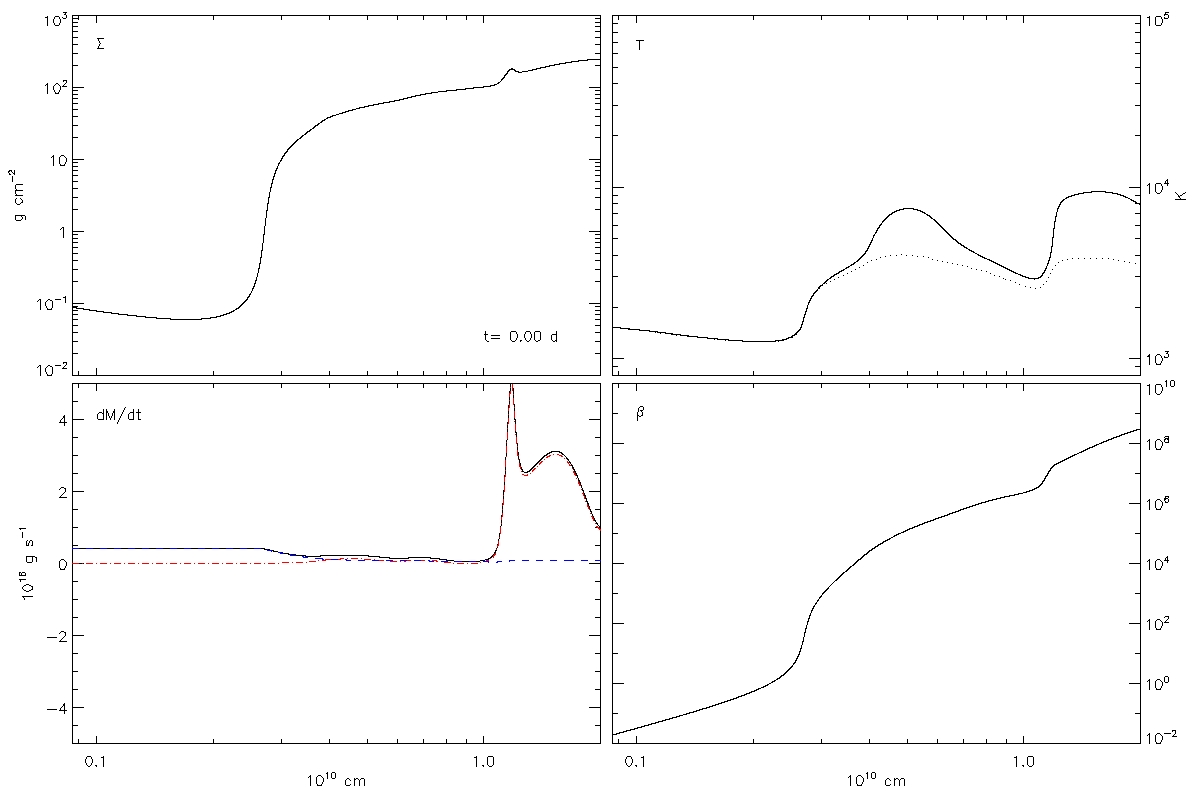}}{VPlayer.swf}
 \caption{(Movie) Time evolution of the radial structure of the unstable disk with $\mu_{30}=1$. Clockwise from top left: $\Sigma$; $T_{c}$ and $T_{\rm eff}$ (dotted line); $\beta$; $\dot{M}$ (solid line), $\dot{M}_{r\phi}$ (red dot dashed line), and $\dot{M}_{z\phi}$ (blue dashed line). The time $t$ in the $\Sigma$ panel corresponds to the time in the $\mu_{30}=1$ light curve shown in  Fig.~\ref{fig:lcdipole}. Requires Adobe Reader.\label{fig:movie}}
\end{figure*}

The situation is completely different in the dipolar case. The lower panels of Fig.\,\ref{fig:lcdipole} are reminiscent of the light curves of dwarf novae.  Actually, the bottom light curve has outbursts with an amplitude in $V$ and a recurrence time comparable to U Gem. An increasing $\mu$ leads to a greater region of the inner disk becoming wind-dominated (Eq.\,\ref{eq:rmu}). Again, the wind torque resembles a truncation of the inner disk. The critical $\Sigma$ required to trigger an outburst increases with radius ($\propto r^{1.1}$, \citealt{1998MNRAS.298.1048H}), so that a truncated disk needs to build up more mass between outbursts, increasing the quiescence time.  However, the impact of the magnetic wind torque on the light curve cannot be reduced to this truncation. Running a disk model with an inner disk truncated at $r_{\mu}$ and leaving out the wind torque produces weak outbursts with short cycles. Including the wind torque is critical because it dominates during the decline from outburst (Fig.\,\ref{fig:movie}), imposing a higher mass accretion rate, with more of the disk mass accreted. This leads to larger outburst amplitude and longer recurrence time than models without wind torque.

For $\mu_{30}\ga 3$ the disk ends up cold and stable (staying in `quiescence'). The radial structure then becomes similar to those shown in Fig.~\ref{fig:stabhotdipole} with a wind-dominated inner disk and a cold, viscous outer disk (instead of a hot, viscous outer disk). Like magnetospheric truncation, a wind torque can stabilize an unstable disk. It does this for much lower values of $\mu$ (or higher values of $\dot{M}_t$) than magnetospheric truncation would predict because $r_\mu \approx 2.8 r_A$ (Eqs.\,\ref{eq:rmu} and \ref{eq:trunc}).

Figure~\ref{fig:movie} shows the time evolution of the radial structure of the disk with $\mu_{30}=1$, corresponding to the third light curve in Fig.\,\ref{fig:lcdipole}. The movie starts at the beginning of quiescence, $t\approx 0.0\rm\,d$ in Fig.\,\ref{fig:lcdipole}. Mass diffuses inward until the appearance of two heat fronts at $t\approx 9.4\rm\,d$ signals the start of the outburst. The disk is then heated by the outward-moving front, and the wind-dominated region gradually shrinks as $\dot{M}$ increases. The outer radius of the wind-dominated region, numerically traced by the location where $\beta=100$, is within 20\% of the radius $r_{\mu}$ calculated using Eq.\,\ref{eq:rmu}.

\subsection{Conditions for disk stability\label{sec:stab}}
\begin{figure}
\begin{center}
\includegraphics[width=\linewidth]{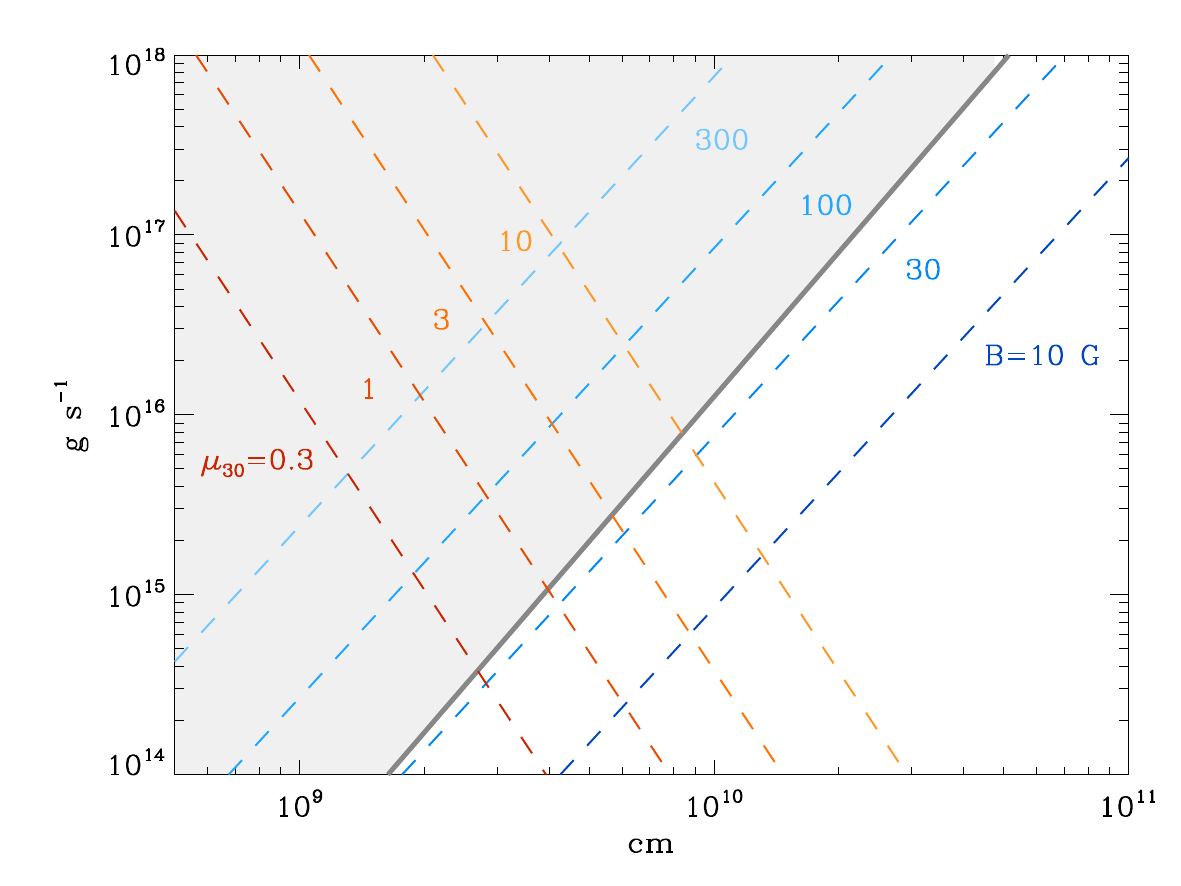} 
\caption{Disk stability in the $(R_{\rm D},\dot{M})$ plane. The solid line is the critical $\dot{M}(R)$ separating stable disks (shaded region) from  disks  that are unstable to the thermal-viscous instability. The dashed lines show $r_{B}$ (shades of blue) and $r_{\mu}$ (shades of red) for various values of the net magnetic field.}
\label{fig:stab}
\end{center}
\end{figure}
Figure~\ref{fig:stab} is useful in order to anticipate the impact of the wind torque on the stability of the accretion disk. The critical mass accretion rate above which a purely viscous disk of radius $r$ is stable can be approximated as \citep{1998MNRAS.298.1048H}
\begin{equation}
 \dot{M}_{\rm crit}=8.0\times 10^{15}\ r_{10}^{2.67} M_{1}^{-0.89}\rm\,g\,s^{-1}.
 \label{eq:crit}
\end{equation}
For example, taking $R_{\rm d}=2\times 10^{10}\rm\,cm$ and $\dot{M}=3\times 10^{17}\rm\,g\,s^{-1}$, the location of the disk in Fig.\,\ref{fig:stab} shows that it is in the stable region and that a wind-dominated region appears in the outer disk for a constant $B> 70\rm\,G$, in agreement with the results of Fig.\,\ref{fig:stabhotcomp}. Similarly, as can be seen from the plotted $r_{\mu}$ lines  in Fig.\,\ref{fig:stab}, a wind-dominated inner region appears for $\mu_{30}>1$, which is consistent with Fig.\,\ref{fig:stabhotdipole}. The disk becomes fully wind-dominated if either $\mu_{30}\ga 280$ (using Eq.\,\ref{eq:rb}) or $B\ga 3\,500\rm\,G$ (Eq.\,\ref{eq:rmu} with $R_{\rm in}=8.7\times 10^{8}\rm\,cm$). Such a disk would accrete at a very high rate yet emit very little radiation since wind-driven transport dissipates no energy in the disk.

Lowering $\dot{M}$ to $10^{16}\rm\,g\,s^{-1}$, the disk is now in the unstable region of the diagram. The disk is truncated by a wind-dominated region when $B=30\rm\,G$ and is nearly stable according to Fig.\,\ref{fig:lccomp}, in agreement with Fig.\,\ref{fig:stab}.  Above 30\,G, the disk becomes hot and stable, with the wind torque dominating in the outer disk. Similarly, we recover the results of Fig.\,\ref{fig:lccomp}, with the disk becoming cold and stable if $r_{\mu}\ga r_{\rm crit}$ (with $r_{\rm crit}$ defined from Eq.\,\ref{eq:crit}). However, in this case, this criterion overestimates the required $\mu$ to stabilize the disk: we find $\mu_{30}\ga 3$ stabilizes the disk from Fig.\,\ref{fig:lcdipole} whereas the criterion gives $\mu_{30}\ga 10$. Even if the wind torque does not dominate, the density and temperature are lower at larger radii than in a viscous disk, stabilizing the disk in the cold state earlier than expected from the simple approximation.

\section{Discussion\label{sec:disc}}
Our results reveal that including a magnetic wind torque, under the form derived from MRI shearing box simulations, strongly impacts the dynamics of dwarf novae accretion disks. Regions with $\beta\la 10^4$ experience a runaway increase in the wind torque, resulting in a cold disk with a high accretion rate. These wind-dominated regions have a short accretion timescale, such that $\dot{M}$ is close to constant with $r$. Thus, they behave like a  truncation of the disk in terms of its dynamics. However, the impact of wind torque does not reduce to a truncation; it also contributes to the mass flow rate during the outburst cycle, changing the light curves dramatically. This transition is reminiscent of the jet-emitting disk/standard accretion disk transition proposed in the context of X-ray binaries in \cite{ferreira2006}.

\subsection{Signatures of wind-dominated regions}
When assuming a dipolar configuration for $B_z$ we find light curves that have outburst amplitude in $V$, outburst duration and recurrence time in total agreement with observed DNe. Relatively weak values of the dipolar moment  $\mu\sim 10^{30}\rm\,G\,cm^3$ are sufficient to obtain those light curves. These values of $\mu$ are consistent with those expected of dwarf novae, and are two orders of magnitude weaker than those in intermediate polars where the disk is truncated by the white dwarf magnetosphere; in the $B_z$ configuration we explored, the wind torque will always truncate the inner disk before the magnetosphere does. There are two main observable consequences. First, the mass accretion rate onto the white dwarf in quiescence is higher than would be expected from a purely viscous disk, in better agreement with the high X-ray luminosities measured in quiescence (\citealt{wheatley2003}, \citealt{collins2010}, \citealt{mukai2017}). Magnetospheric accretion or coronal evaporation have also been proposed to explain these X-ray luminosities \citep{lasota2001}, playing on the truncation of the inner disk to obtain higher mass accretion rates in quiescence. However, in our model, accretion would still proceed in a wind-dominated disk all the way to the white dwarf surface, and we expect the X-ray emission to arise from the boundary layer, in better agreement with observations of quiescent DNe \citep{mukai2017}. 

\subsection{Dark accretion}
The wind-dominated regions also stabilize the disk if they are large enough to leave the disk with only one hot (or cold) viscous region. There is no heating in the wind-dominated region so the disk will appear fainter than its purely viscous equivalent, as was previously noted by \cite{ferreira1995}. Peculiar observational signatures may trace these disks. For instance, a smaller disk than expected given the system's orbital parameters, with the discrepancy showing up from the disk eclipse by the secondary or from the double-peaked profile of the disk lines, for example. The whole disk can also become wind-dominated if $\beta$ is everywhere $\la 10^4$ (Sect.\,\ref{sec:stab}). The only indication for the presence of such a dark accretion disk would  be a high X-ray or UV luminosity from the boundary layer and radiation from the impact of the Roche lobe overflow stream with the outer disk. However, we note that X-ray irradiation and reprocessing might increase the luminosity coming from these dark disks and produce an observational signature such as an Fe line.

\subsection{Impact on the predictions of the DIM}
In principle, the wind torque can stabilise a disk anywhere in the $(R_{d},\dot{M}_{t})$ plane (Fig.\,\ref{fig:stab}), going against the standard DIM predictions. On the one hand, this is likely to be rare or \citet{2018A&A...617A..26D} would not have found very good agreement between the distribution of systems in this plane and their stability properties. On the other hand, perhaps this results from an observational bias against detecting faint stable cataclysmic variables. Among the bright cataloged systems, \citet{2018A&A...617A..26D} found that AE Aqr is the only stable cataclysmic variable with an inferred $\dot{M}_{t}$ that puts it in the unstable region. AE Aqr is an intermediate polar with one of the fastest  spinning white dwarfs known, truncating the inner disk and removing part of the expected optical flux from the disk. In this case, using the optical magnitude with a full disk underestimates the true $\dot{M}_{t}$ \citep{2018A&A...617A..26D}. The same effect is expected if a significant wind torque is at play. For example, the average $\dot{M}$ reconstructed from the $V$-band light curves in Fig\,\ref{fig:lcdipole}, following the method of \citet{2018A&A...617A..26D}, are lower than the true $\dot{M}_{t}$ by factors of 0.7, 0.4, 0.2, and 0.1 for $\mu_{30}=0.3, 1, 2$, and 2.5, respectively. This might explain some of the dispersion in $\dot{M}$ around the secular mass transfer rate expected for a given orbital period (see Fig.\,3 in \citealt{2018A&A...617A..26D}) as variations from system to system in the distribution of $B_{z}$.

\section{Conclusion}
A model of dwarf novae light curves based upon the first principles of physics remains elusive, yet these systems continue to provide an unrivaled probe into the properties of transport in accretion disks. The transport coefficients derived from zero net flux simulations fail to produce realistic looking light curves when incorporated in the disk evolution equations \citep{2016MNRAS.462.3710C}. But here we find that including a net magnetic flux, resulting in an additional magnetic torque on the disk, may be the key to getting light curves with the right properties (outburst amplitude, duration, and recurrence time). However, this will not solve the vexing issue of heating in quiescence since a net magnetic flux does not help maintain MRI at low ionization fractions \citep{2018arXiv180909131S}.

Our results obviously depend on the radial and temporal distribution of the net $B_{z}$ in the disk. The light curves favor a distribution leading to an increase in $\beta$ outwards, {\em i.e.}, a higher net magnetic flux at small radii. Whether these configurations arise depends on the origin of the vertical magnetic field and its transport in the accretion disk with time. It will be crucial to check whether including the advection and diffusion of $B_{z}$ in the disk will still yield light curves that look like dwarf novae. Future work should also account for the mass lost in the magnetic wind, and have a better description of the low $\beta$ disks. In the end, the light curves of accreting systems may be more revealing of the magnetic field threading the disk than of the $\alpha$ coefficient of turbulent transport.

The DIM is also at work in X-ray binaries, whose disks sample  density and temperature conditions comparable to those of DNe \citep{lasota2001}.  Magnetic flux concentrations, demultiplied by the high ratio of outer to inner radii in X-ray binaries compared to DNe, would similarly impact the dynamics of the disk through the magnetic wind torque, perhaps explaining some of the bewildering range of X-ray binary light curves. Enhanced accretion due to the wind torque could  thus explain the fast rates of decline from outbursts, which require unrealistically high values of $\alpha \ga 1$ with a purely viscous disk \citep{2018Natur.554...69T}. The runaway to $\beta\approx 1$ in wind-dominated regions could  also result in the formation of the magnetically dominated jet-emitting disk proposed by \cite{ferreira2006}, which can account for the changes in radio and X-ray emission during the outbursts of GX 339-4 \citep{2018A&A...617A..46M}.  It had been speculated that high concentrations of magnetic flux in the inner regions are the reason for emission state changes in X-ray binaries \citep{2008MNRAS.385L..88P,2014ApJ...782L..18B,2015ApJ...809..118B,2016ApJ...817...71C}. However, low-mass X-ray binaries seem to lose an important fraction of their disk mass to a wind \citep{ponti2012}. Whether these winds are thermally and/or magnetically launched is still under debate \citep{luketic2010,higginbottom2015,miller2006,chakravorty2016,diaz2016}. In any case, wind mass loss will be an important factor to include in future applications of our model of these systems.

\begin{acknowledgement}
We thank the referee for his comments and suggestions. The authors also wish to thank Jonathan Ferreira, Pierre-Olivier Petrucci and Jean-Pierre Lasota-Hirszowicz for the very fruitful discussions. NS acknowledges financial support from the pole PAGE of the Universit\'e Grenoble Alpes. This work was granted access to the HPC resources of IDRIS under the allocation A0040402231 made by GENCI (Grand Equipement National de Calcul Intensif). Some of the computations presented in this paper were performed using the Froggy platform of the CIMENT infrastructure (https://ciment.ujf-grenoble.fr), which is supported by the Rhône-Alpes region (GRANT CPER07\_13 CIRA), the OSUG@2020 labex (reference ANR10 LABX56) and the Equip@Meso project (reference ANR-10-EQPX-29-01) of the programme Investissements d'Avenir supervised by the Agence Nationale pour la Recherche. 
\end{acknowledgement}

\bibliographystyle{aa}
\bibliography{MRIwind}


\end{document}